# Physicochemical influences on electrohydrodynamic transport in compressible packed beds of colloidal boehmite particles

Bastian Schäfer *, Hermann Nirschl

Dipl.-Ing. B- Schäfer *, bastian.schaefer@mvm.uka.de,
Tel. ++497216082411, Fax. ++497216082403
Prof. Dr.-Ing. H. Nirschl, hermann.nirschl@mvm.uka.de
Tel. ++497216082401, Fax. ++497216082403

Institut für Mechanische Verfahrenstechnik und Mechanik
Universität Karlsruhe (TH), Straße am Forum 8, 76131 Karlsruhe, Deutschland

# 1 Abstract

Production and processing of colloidal particles require a deeper understanding of surface charge of particles and interaction of mass and charge transport in packed beds. The assessment of fundamental parameters is rather complex due to the additional influence of the particle charge on the structure of a packed bed. The combination of different measurement techniques (streaming potential and electroosmosis) allows for separating the effects, based on the postulation of a new method to quantify the ratio of surface conductance to liquid conductance. The purpose of this paper is to investigate the influence of the pH value and compression on the electrohydrodynamic transport parameters.

## 2 Introduction

In a packed bed single colloidal particles are randomly and densely distributed between two membranes while liquid flows through the porous structure. Understanding the influences of physico-chemistry on the mass and charge transport in packed beds is important for industrial applications: Colloidal particles have promising applications, but the poor permeability of filter cakes limits mechanical separation and hence production and processing [1]. The consolidation of green bodies is essential for realizing transparent nanoceramics. Nanoporous packed beds could also be used in fixed bed catalyst reactors and electroosmotic micropumps. Porous ceramic bodies are used as a barrier between the anode and the cathode in lithium ion batteries.

The problem of the very low permeability can be overcome by influencing the agglomerate structure of the particles or by utilizing the interaction of hydraulic and electric transport. Therefore the focus of this paper is to understand the combination of electrohydrodynamic transport phenomena and electrostatic effects on the packed bed structure [2] and to separate them from each other.

In this study, the influence of the pH value and the compression on the transport through a nanoporous packed bed is investigated. The packed beds are formed by filtration and compression of colloidal boehmite suspensions. A new method is developed for combining different measurement techniques in order to separate the interrelated influences. The method allows for determining the surface charge and the electrohydrodynamic transport in packed beds. The experimental conditions are well in the range of the thin electrochemical double layer (EDL).

## 3 Theoretical background

Electrohydrodynamic transport phenomena in nanoporous packed beds are closely related to the particle charge and its influence on the pore structure of packed beds. The EDL on ceramic particles in aqueous suspensions originates from dissociation reactions on the particle surface atoms (denoted by $S$) [3, 4]. The equilibrium of the dissociation depends on the concentration of hydroxide ions:

$$SOH \rightleftharpoons SO^- + H^+,$$
$$SOH + H_2O \rightleftharpoons SOH_2^+ + OH^-.$$

[1]

The material specific pH value for which the surface charge equals zero is called isoelectric point (IEP). Below the IEP the particle charge is positive and above the IEP it becomes negative. The particle is surrounded by a diffuse cloud of counter-ions, which are attracted to the particle by electrostatic forces. The exponential decay length $\kappa^{-1}$ of the ion concentration, the Debye length, is a function of the ionic strength $\tilde{I}$ with the Faraday constant $F$, the relative permittivity of the liquid $\varepsilon_L$, the vacuum permittivity $\varepsilon_0$, the gas constant $R$ and the temperature $T$ [3, 4]:

$$\kappa^{-1} = \left(\frac{2F^2\tilde{I}}{\varepsilon_L\varepsilon_0 RT}\right)^{-\frac{1}{2}}. \qquad [2]$$

Equ. 2 was changed because the exponent on the Faraday constant was missing.

In our study, at $\tilde{I} = 0.01$ M and $T = 288$ K, the Debye length is approximately 3 nm. Pragmatically, the EDL is subdivided into the immobile Stern layer (SL) and the surrounding diffuse layer (DL). As the surface charge density is difficult to measure, the EDL is often characterized by the zeta potential $\zeta$, which is more relevant for practise and easier to determine, e.g. by analyzing the electrokinetic effects [3, 5]. The latter are described by cross coefficients in the electrohydrodynamic transport equations, which relate the volumetric flow $\dot{V}$ and electric current $I$ to the difference of the pressure $p$ and the electric potential $\Psi_{Ext}$:

$$\dot{V} = C_{11}\Delta(-p) + C_{12}\Delta(-\Psi_{Ext}),$$
$$I = C_{21}\Delta(-p) + C_{22}\Delta(-\Psi_{Ext}). \qquad [3]$$

$C_{11}$ to $C_{22}$ are material specific coefficients. $C_{12}$ equals $C_{21}$ according to Onsager's relation [6].

- $C_{11}$ is related to Darcy's law, which describes the flow rate $\dot{V}$ of a liquid with the viscosity $\eta_L$ driven by the pressure difference $\Delta(-p)$ depending on cross section area $A_{PB}$ and the height $L_{PB}$ of a packed bed [7]:

$$\dot{V} = \frac{K_{hydr} A_{PB}}{\eta_L L_{PB}} \Delta(-p). \qquad [4]$$

Several models for the permeability $K_{hydr}$ have been proposed for macroporous packed beds with a homogenous pore size, e.g. by Carman and Kozeny [8]. Briefly, an increase of porosity and particle size leads to a higher permeability. In case of a compressible packed bed, an increase of compression leads to a decrease of porosity [9]. As the liquid permeates the structure, the liquid pressure is reduced due to friction and transmitted to the solid structure. Consequently, the compression has a maximum and the porosity has a minimum at the downstream membrane [9]. In this study, the effect of an inhomogeneous compression can be neglected as the packed bed is consolidated evenly by a plunger. For nanoporous packed beds, the porosity and the pore size distribution depend largely on the electrostatic effect, i.e. the agglomerate structure [2]. Agglomeration is determined by the equilibrium of electrostatic repulsion, van-der-Waals attraction and Born's repulsion between particles as described by the DLVO theory. It can be investigated by small-angle x-ray scattering [3, 4, 10]. Suspensions with a low particle charge tend to agglomerate, leading to a loosely structured packed bed with an inhomogeneous pore size distribution. The large pores between the agglomerates are accountable for the high permeability. In contrast, a stable suspension with a high zeta potential leads to a dense structure with a homogenous pore size distribution and a low permeability [2].

- $C_{12}$ describes the electroosmotic mass transport, driven by an external electric potential $\Psi_{Ext}$. The counter-ions in the DL of the particles are accelerated and drag the adjacent water molecules along. Smoluchowski's assumption of a thin double-layer, or more precisely $\kappa a \gg 1$ with the reciprocal Debye length $\kappa$ and the radius curvature $a$, is fulfilled here. In this case, the electroosmotic flow $\dot{V}_{EO}$ in a hypothetical cylindrical capillary depends on the zeta potential and the cross section area $A_{Cap}$ [3, 5]:

$$\dot{V}_{EO} = -\frac{\varepsilon_L \varepsilon_0 \zeta}{\eta_L} A_{Cap} \cdot \Delta(-\Psi_{Ext}). \qquad [5]$$

The electric potential $\Delta(-\Psi_{Ext})$ is the quotient of the externally applied voltage $U_{Ext}$ and the length of the capillary $L_{Cap}$. Experiments showed that the complex pore geometry of a packed bed with the cross section area $A_{PB}$, the thickness $L_{PB}$ and the porosity $\Phi_L$ can be accounted for by substituting $A_{Cap}/L_{Cap}$ with $A_{PB}/L_{PB} \cdot \phi_L^{2.5}$ [5, 11]:

$$\dot{V}_{EO} = -\frac{\varepsilon_L \varepsilon_0 \zeta}{\eta_L} \frac{A_{Cap}}{L_{Cap}} U_{Ext} = -\frac{\varepsilon_L \varepsilon_0 \zeta}{\eta_L} \frac{A_{PB}}{L_{PB}} \phi_L^{2.5} U_{Ext}. \qquad [6]$$

- $C_{21}$ specifies the streaming current $I_{Str}$, which is the charge transport generated when counter-ions are dragged from the DL by a pressure-driven liquid-flow. The streaming current can be measured if the inlet face of the structure is connected to the outlet face with a low-resistance ampere meter (short-circuit conditions) [5]:

$$I_{Str} = -\frac{\varepsilon_L \varepsilon_0 \zeta}{\eta_L} \frac{A_{Cap}}{L_{Cap}} \Delta(-p) = -\frac{\varepsilon_L \varepsilon_0 \zeta}{\eta_L} \frac{A_{PB}}{L_{PB}} \phi_L^{2.5} \Delta(-p). \qquad [7]$$

- C22 is the electric conductance of the porous structure. The conductivity K comprises liquid conduction $K_L$ (ion movement in the electrolyte solution in the pores) and surface conduction $K_{Surf}$ (movement of excess ions in the DL and SL). Surface conductance causes a "short-circuiting" inside the packed bed, which must not be neglected for a correct interpretation of electrohydrodynamics. For a cylindrical capillary with the radius $r_{Cap}$, geometrical considerations leads to

$$K = K_L + 2K_{Surf}/r_{Cap} = K_L(1 + 2Du) \qquad [8]$$

with the Dukhin number $Du$ [5, 12]:

$$Du = K_{Surf}/(K_L r_{Cap}) = \frac{1}{2}\left(\frac{K}{K_L} - 1\right). \qquad [9]$$

The total conductivity of the sample is calculated as $K = (L/A)/R$, with the geometry factor $L/A$ and the electric resistance $R$. For a non-deformable structure the geometry factor can be determined from the resistance $R^\infty$ when the structure is filled with high ionic strength electrolyte solution with the conductivity $K_L^\infty$, so that surface conduction is negligible:

$$L/A = K_L^\infty R^\infty. \qquad [10]$$

However, the technique is not applicable to packed beds since the increase of ionic strength influences the packed bed's structure significantly. We propose to derive the Dukhin number from Eq. [8] with $\zeta$ from Eq. [6] and the ratio of the electroosmotic flow to the electric current $I_{Ext}$ as given by [5]

$$\frac{\dot{V}_{EO}}{I_{Ext}} = -\frac{\varepsilon_L \varepsilon_0 \zeta}{\eta_L} \frac{R_{PB}}{K_L^\infty R^\infty}, \qquad [11]$$

$$Du = -\frac{1}{2}\left(\frac{\varepsilon_L \varepsilon_0 \zeta}{\eta K_L \frac{\dot{V}_{EO}}{I_{Ext}}} + 1\right). \qquad [12]$$

Our approach is analogous to a method introduced by [13], but is not limited to straight capillaries.

- If counter-ions are dragged from the DL by a pressure-driven liquid-flow and the faces of the packed bed are not short-circuited, the streaming current has to be balanced. Hence, the streaming potential $U_{Str}$ is built up, which drives counter-ions back against the pressure driven flow. $U_{Str}$ is measured between an upstream and a downstream electrode and can be used to determine $\zeta$ if surface conduction is considered. Surface conduction reduces the potential required to balance the streaming current [3, 5]:

$$U_{Str} = \frac{\varepsilon_L \varepsilon_0 \zeta}{\eta_L} \frac{R_{PB}}{K_L^\infty R_{PB}^\infty} \Delta(-p). \qquad [13]$$

- The counter-ions driven against the flow by the streaming potential induce an electroosmotic backflow, which retards the permeation. This apparent rise of the fluid viscosity, the so called electroviscous retardation, can be calculated if the electric current $I$ in Eq. [3] is set to zero [14]:

$$\dot{V} = \left(C_{11} - \frac{C_{12}C_{21}}{C_{22}}\right)\Delta(-p). \qquad [14]$$

We define the electroviscosity ratio $EVR$ as the ratio between the apparent viscosity $\eta_{App}$ and the real viscosity of the permeate $\eta_L$:

$$EVR = \frac{\eta_{App} - \eta_L}{\eta_L} = \frac{C_{12}C_{21}}{C_{11}.C_{22} - C_{12}^2}. \qquad [15]$$

The $EVR$ allows for quantifying the electroviscous retardation.

## 4  Experiments

Packed beds are formed in the cylindrical Electro-Compression-Permeability-Cell (ECP-Cell) by filtration of 200g suspension. The packed bed is then compressed by a plunger and permeated by a voltage of a pressure difference. The two-sided filtration between two nylon membranes (Ultipor from Pall, USA, with a nominal pore size of 0.1µm) in the vertical position of the ECP-Cell and the low thickness-to-diameter ratio ensure a constant compression and thus homogenous structure of the packed bed. All parts of the ECP-Cell except for the electrodes are nonconducting. After the filtrate flow abates, the ECP-Cell is turned to the horizontal position, which is shown in Fig. 1.

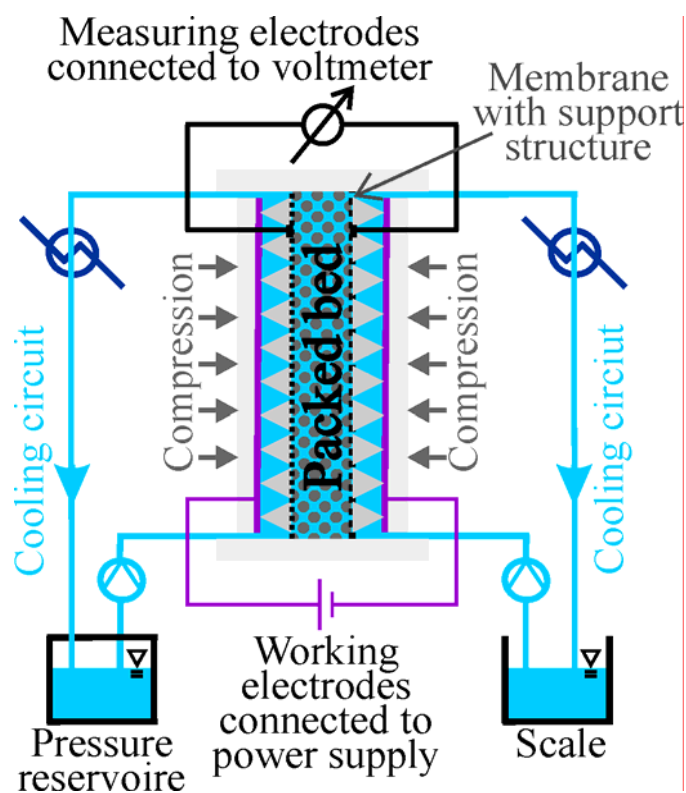

**Figure 1: Electro-Compression-Permeability-Cell in horizontal position**

The packed bed is cooled by circulating electrolyte solution through heat exchangers. The solution has the same ionic strength and pH value as the suspension. After reaching a constant temperature of 290K, the circulation stops and the packed bed is permeated by the electrolyte solution, driven by a pressure difference or by an electric field. The scale displays the accumulation of mass on the downstream side.

The streaming potential evoked by a pressure driven permeation is measured with platinum electrodes (measuring electrodes). Alternatively, an electric field applied via the working electrodes excites electroosmotic permeation. The iridium oxide coating of the titanium electrodes prevents oxidation. If the electric field is reversed every 30 seconds, the formation of electrolytic gas is prevented because hydrogen formed at the cathode reacts with oxygen from the previous cycle. The left working electrode is defined as anode, the right one is the cathode. A negative electric field means that voltage and flow are reversed. The effective voltage at the packed bed is monitored with the measuring electrodes because the voltage is reduced due to the electric resistance of the water in the cooling circuit.

The suspensions are prepared by dispersing 25g boehmite particles (Disperal 40 from Sasol, Germany, with a particle size of 350 nm) in 225g electrolyte solution with a propeller stirrer for 15 minutes. The electrolyte solution has an ionic strength of 0.01 M and a pH value of 4, adjusted with $KNO_3$ and $HNO_3$. While for most particles an elaborate method for dispersing is required [15, 16, 17], Boehmite particles are easily dispersed in acidic solutions, which was verified with a Coulter N4 Plus. Subsequently, the suspension is flocculated by injecting different amounts of 1M KOH into the suspension and stirred for further 30 minutes to homogenize the suspension. The background electrolyte $KNO_3$ guarantees the well-definition of the surface charge [5]. The zeta potential of the suspensions was determined with an AcoustopSizer II (see Fig. 5). The primary particle size distribution is close to a Gaussian distribution with $q_3 = 31,3 \cdot exp\left(-0,5\left(\frac{x-324,6}{60,6}\right)^2\right)$. Boehmite particles are common as filler in synthetics and raw material for the aluminium oxide industry. Boehmite is dissolvable in aqueous solutions, especially at high and low pH values. However, at temperatures below 50°C the kinetics are so slow that the solubility can be neglected [18].

## 5 Results

For the interpretation of the measurements, a homogenous compression of the packed bed and thus a homogenous porosity are assumed because the compression from the plunger is significantly higher than the pressure drop of the permeate. The porosity was calculated from the height of the packed bed $L_{PB}$ and the mass of the particles $m_S$:

$$\phi_L = \frac{\left(L_{PB}A_{PB} - \frac{m_S}{\rho_S}\right)}{L_{PB}A_{PB}}. \qquad [16]$$

In all figures, the error bars represent the standard deviation.

A higher compression leads to a lower porosity (see Fig. 2). The small reduction of the porosity (from 0.8 to 0.72) goes along with a strong decrease of the permeability, since mainly the large pores between the agglomerates, which are accountable for the major part of the fluid transport, are affected [2, 19].

Porosity also changes with a variation of pH value (see Fig. 3) due to the agglomeration of the particles when the pH value approaches the IEP at pH 9.5 (compare Fig. 5). Beyond the IEP, the porosity remains high because of the sample preparation process: The particles can only be dispersed at a low pH value. In order to obtain a high pH value, the IEP has to be passed. The agglomerates formed at the IEP are not disagglomerated thereafter because the shear gradients caused by the stirrer are too small. Again, the increase of the porosity causes an increased permeability at the IEP and beyond it.

The permeability is identical for points of the same porosity when comparing Fig. 2 and Fig. 3. A packed bed from an agglomerated suspension at high compression and a packed bed from an unagglomerated suspension at low compression have the same permeability if the porosity is identical. Accordingly, the change of permeability upon variation of the pH value is attributed to a change of porosity and there is no further effect of the agglomerate structure of the packed bed on the permeability.

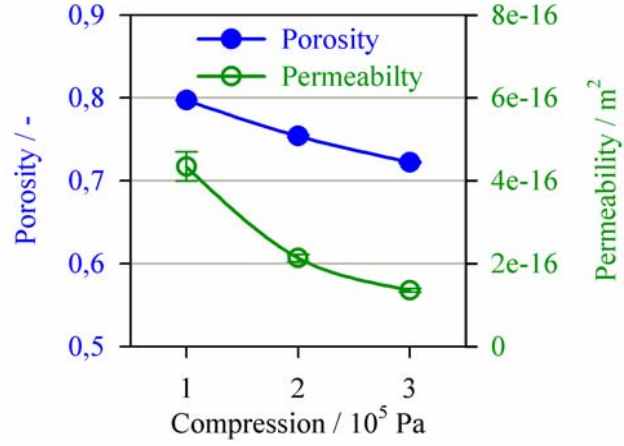

**Figure 2: Porosity and permeability depending on compression for packed beds with a pH value of 10.9 and an ionic strength of 0.01M.**

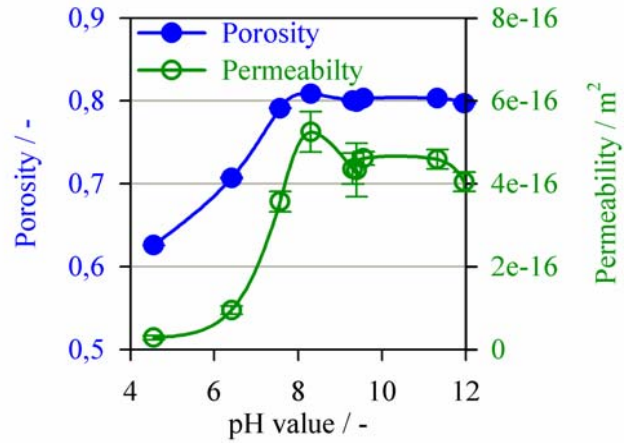

**Figure 3: Porosity and permeability depending on pH value for packed beds with a compression of 100 kPa and an ionic strength of 0.01M.**

For thin double layers ($\kappa a \gg 1$), $\zeta$ is not a function of the compression (see Fig. 4). The zeta potential is not affected by the reduction of the pore size because the EDL of neighbouring particles do not overlap. $\zeta_{\dot{V}_{EO}/\Delta(-\psi)}$ is calculated with Eq. [6] and $\zeta_{U_{Str}/\Delta(-P)}$ with Eq. [13]. $R_{PB}/(K_L^\infty R_{PB}^\infty)$ is determined from the Dukhin number (see below).

Our zeta potential measurements in the packed bed are consistent with those made in suspensions of 10%wt in an AcoustoSizerII (see Fig. 5). $\zeta_{EA,Smol}$ is based on the assumption of the Smoluchowski dynamic electroacoustic limit and $\zeta_{EA,O'Brien}$ was determined with the O'Brien algorithm. The discrepancy between $\zeta_{\dot{V}_{EO}/\Delta(-\psi)}$, $\zeta_{U_{Str}/\Delta(-P)}$ and the AcoustoSizer measurements is in the same order of magnitude as the discrepancy in between the AcoustoSizer measurements. This was expected, as Fig. 4 shows that the solid concentration has no influence on $\zeta$ in the limit of thin double layers. Therefore we can conclude that our measurements techniques and our approach to the Dukhin number are valuable.

For the experiments at a pH value of 10.9, the ratio of the electroosmotic flow to the externally applied electric current $\dot{V}_{EO}/I_{Ext}$ is $6\cdot10^{-8}$ m$^3$/A. This corresponds to $\dot{V}/I \times N_A \times r/M = 2\cdot10^{21}$ water molecules per Coulomb or $\dot{V}/I \times F \times r/M = 318$ water molecules dragged by each counter-ion with the Avogadro constant $N_A$ and the Faraday constant $F$. $\dot{V}_{EO}/I_{Ext}$ does not depend on the compression and thus pore size as the acceleration in the EDL causes the plug profile of electroosmotic flow, so that the velocity is independent of the pore size [3]. $\dot{V}_{EO}/I_{Ext}$ has the same tendency as $\zeta$ when the pH value is varied.

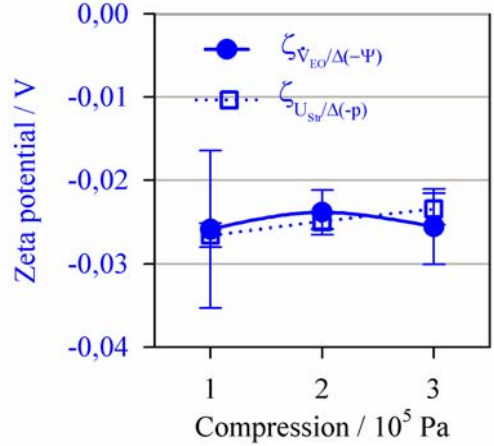

**Figure 4: Zeta potential depending on compression for packed beds with a pH value of 10.9 and an ionic strength of 0.01M. $\zeta_{\dot{V}_{EO}/\Delta(-\psi)}$ is calculated from the electroosmotic flow and $\zeta_{U_{Str}/\Delta(-P)}$ is determined from the streaming potential.**

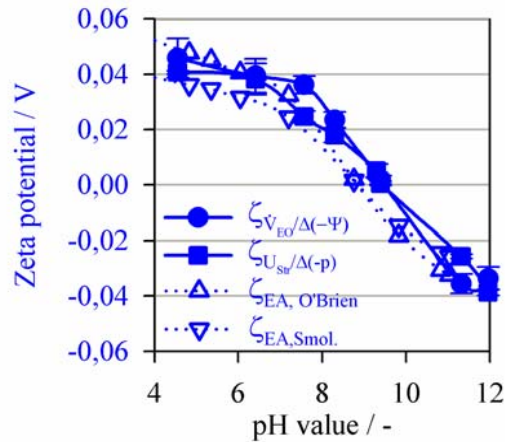

**Figure 5: $\zeta$ depending on pH value to electric current for packed beds with a compression of 100 kPa and an ionic strength of 0.01M. $\zeta_{\dot{V}_{EO}/\Delta(-\psi)}$ was calculated from the electroosmotic flow and $\zeta_{U_{Str}/\Delta(-P)}$ was determined from the streaming potential. $\zeta$EA,Smol and $\zeta$EA,O'Brien are calculated with an algorithm assuming the fulfilment of the Smoluchowski dynamic electroacoustic limit and the OBrien algorithm, from measurements with an AcoustosizerII in a suspension of 10%wt.**

The Dukhin number ($Du$) is calculated by using Eq. [12] with $\zeta_{\dot{V}_{EO}/\Delta(-\psi)}$ from Fig. 5 and the measured ratio of the electroosmotic flow to the externally applied electric current. For the experiments at a pH value of 10.9, a Dukhin number of 0.23 was found, which is independent of the compression. This is reasonable for thin double layers as the reduction of pore size does not influence the surface conduction. The compression is expected to influence the Dukhin number only when the thin double layer approximation is not valid.

In contrast, the influence of the pH value on the Dukhin number is fitted with a quadratic function (see Fig. 6). The minimum of the Dukhin number at the IEP is caused by the low zeta potential: The weak EDL does not cause an excess surface conduction, which can also be seen from the maximum of the electric resistance. The reduced volume available for charge transport may be the reason for the negative value of *Du*. But as both $\zeta$ and $\dot{V}_{EO}/I$ in Eq. [12] are close to zero at the IEP, the absolute value and the negative sign are rather uncertain.

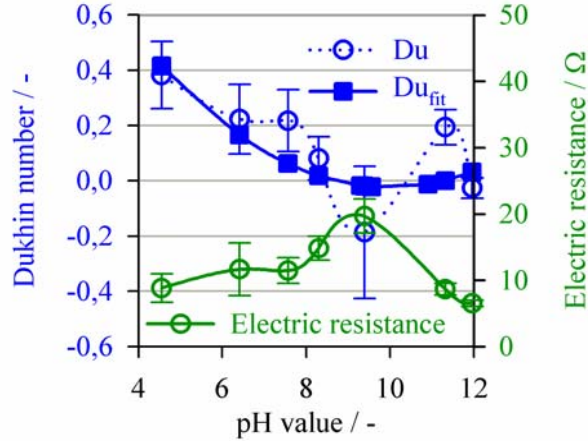

**Figure 6: Dukhin number (*Du*), fit of the Dukhin number (*Du$_{fit}$*) and electric resistance depending on the pH value for packed beds with a compression of 100 kPa and an ionic strength of 0.01M.**

The electroviscosity ratio *EVR* increases with increasing compression (see table 1), because a larger part of the flow takes place inside of the EDL when the pores get smaller. The electroviscosity ratio is zero at the IEP (see Fig. 7), where no streaming potential is build up and no electroosmotic backflow evolves. Beyond the IEP, *EVR* is smaller than below the IEP because of the increased porosity due to agglomeration. But altogether, the increase of the apparent viscosity is negligible in the thin double layer limit (i.e. $\kappa a \gg 1$). This corresponds to the finding that the permeability is identical for points of the same porosity when comparing Fig. 2 and Fig. 3. Accordingly, the change of permeability upon variation of the pH value is attributed to a change of porosity, not to the electroviscosity.

**Table 1: Electroviscosity ratio depending on compression for packed beds with an ionic strength of I = 0.01M**

| pH value / - | Compression / Pa | Electroviscosity ratio / - |
|---|---|---|
| 10.9 | 1·10$^5$ | 9.86e-6 |
| 10.9 | 2·10$^5$ | 1.26e-5 |
| 10.9 | 3·10$^5$ | 1.92e-5 |

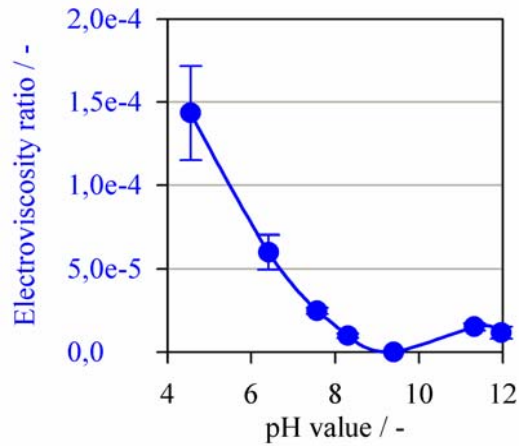

**Figure 7: Electroviscosity ratio depending on pH value for packed beds with a compression of 100 kPa and an ionic strength of 0.01M.**

# 6 Summary

The interrelation of the influence of particle charge on the structure of a packed bed with the interactions of charge and mass transport complicates the understanding of electrohydrodynamic transport phenomena. A combination of different measurement techniques (streaming potential and electroosmosis) allows for separating the effects, based on the postulation of a new method to quantify the ratio of surface conductance to liquid conductance. The influence of pH value and compression is in the focus of this paper.

The zeta potential of the particles in the packed bed depending on pH value is consistent with the zeta potential of suspensions measured with an AcoustoSizerII. $\zeta$ does not depend on the compression of the packed bed in the case of thin double layers (i.e. $\kappa a \gg 1$). Also the ratio of the electroosmotic flow to the electric current is independent of the compression. The ratio of surface conductance to liquid conductance has a minimum at the isoelectric point. The electroviscous retardation of a hydraulic permeation is negligible and change of permeability upon variation of the pH value is attributed to the change of the porosity.

Our approach offers a possibility to access the electrohydrodynamic transport properties of nanoporous materials. Future studies will include the influence of ionic strength and particle size, so that a basis for industrial applications will be established.

# 7 Acknowledgements

We want to thank the German Science Foundation (DFG) for funding within the priority program SPP 1164.